\documentclass[11pt,a4paper]{article}
\usepackage{amsmath,latexsym,fullpage,amssymb,color}
\usepackage{epic,eepic,ifthen,graphics,epsfig}
\usepackage[english]{babel}
\usepackage{times}
\usepackage{float}
\usepackage{xspace}
\usepackage[T1]{fontenc}
\usepackage{tablefootnote}

\newlength {\squarewidth}

\newtheorem{theorem}{Theorem}

\newtheorem{claim}{Claim}

\newcommand{\toto}{xxx}
\newenvironment{proofT}{\noindent{\bf Proof }}
{\hspace*{\fill}$\Box_{Theorem~\ref{\toto}}$\par\vspace{3mm}}

\newenvironment{proofCL}{\noindent{\bf Proof }}
{\hspace*{\fill}$\Box_{Claim~\ref{\toto}}$\par\vspace{3mm}}

\newcounter{linecounter}
\newcommand{\linenumbering}{\ifthenelse{\value{linecounter}<10}
{(\arabic{linecounter})}{(\arabic{linecounter})}}
\renewcommand{\line}[1]{\refstepcounter{linecounter}\label{#1}\linenumbering}
\newcommand{\resetline}[1]{\setcounter{linecounter}{0}#1}
\renewcommand{\thelinecounter}{\ifnum \value{linecounter} >
9 \else \fi\arabic{linecounter}}

\newfloat{algorithm}{thp}{lop}
\floatname{algorithm}{Algorithm}
\newfloat{construction}{thp}{lop}
\floatname{construction}{Construction}
\newcommand{\Xomit}[1]{}

\begin{document}

\title{\bf Optimal Memory-Anonymous \\
           Symmetric  Deadlock-Free Mutual Exclusion}

\author{Zahra  Aghazadeh$^{\dag}$,
        Damien  Imbs$^{\ddag}$, 
        Michel  Raynal$^{\star,\circ}$,
        Gadi    Taubenfeld$^{\diamond}$,
        Philipp Woelfel$^{\dag}$ \\~\\
$^{\dag}$Department of Computer Science, University of Calgary, Canada\\
$^{\ddag}$Aix-Marseille University, CNRS, University of Toulon, LIS, Marseille, France\\
$^{\star}$Univ Rennes IRISA, France \\
$^{\circ}$Department of Computing, Polytechnic University, Hong Kong\\
$^{\diamond}$The Interdisciplinary Center, Herzliya, Israel}

\date{}

\newcommand{\ggcd}{{\sf{gcd}}}
\newcommand{\compareandswap}{{\sf{compare\&swap}}}
\newcommand{\ttrue}{{\tt{true}}}
\newcommand{\ffalse}{{\tt{false}}}

\newcommand{\ssnapshot}{{\sf{snapshot}}}
\newcommand{\wwrite}{{\sf{write}}}
\newcommand{\rread}{{\sf{read}}}
\newcommand{\iisstable}{{\sf{is\_stable}}}
\newcommand{\sshrink}{{\sf{shrink}}}
\newcommand{\rreturn}{{\sf{return}}}
\newcommand{\aacquire}{{\sf{acquire}}}
\newcommand{\rrelease}{{\sf{release}}}
\newcommand{\llock}{{\sf{lock}}}
\newcommand{\uunlock}{{\sf{unlock}}}
\newcommand{\ccount}{{\sf{count}}}
\newcommand{\oowned}{{\sf{owned}}}

  \maketitle

\begin{abstract}
  The notion of an anonymous shared memory (recently introduced in
  PODC 2017) considers that processes
  use different names for the same memory location.  As an example, a
  location name $A$ used by a process $p$ and a location name $B\neq
  A$ used by another process $q$ can correspond to the very same
  memory location $X$, and similarly for the names $B$ used by $p$ and
  $A$ used by $q$ which may (or may not) correspond to the same memory
  location $Y\neq X$.  Hence, there is permanent disagreement on the
  location names among processes. In this context, the PODC paper
  presented --among other results-- a symmetric deadlock-free mutual
  exclusion (mutex) algorithm for two processes and a necessary
  condition on the size $m$ of the anonymous memory for the existence of a
  symmetric deadlock-free mutex algorithm in an $n$-process
  system. This condition states that $m$ must be greater than $1$ and
  belong to the set
  $M(n)=\{m: \forall~ \ell: 1<\ell \leq n:~ \ggcd(\ell,m)=1\}$
  ({\it symmetric} means that, while each process has its own identity,
  process identities can only be compared with equality).

  The present paper answers several open problems related to symmetric
  deadlock-free mutual exclusion in an $n$-process system ($n\geq 2$)
  where the processes communicate through $m$ 
  registers. It first presents two algorithms.  The first considers
  that the registers are anonymous read/write atomic registers and
  works for any $m$ greater than $1$ and belonging to the set
  $M(n)$. Hence, it shows that this condition on $m$ is both necessary
  and sufficient.  The second algorithm considers that the registers
  are anonymous read/modify/write atomic registers. It assumes that
  $m\in M(n)$.  These algorithms differ in their design principles and
  their costs (measured as the number of registers which must contain
  the identity of a process to allow it to enter the critical
  section).  The paper also shows that the condition $m\in M(n)$ is
  necessary for deadlock-free mutex on top of anonymous
  read/modify/write atomic registers.  It follows that, when $m>1$,
  $m\in M(n)$ is a tight characterization of the size of the anonymous
  shared memory needed to solve deadlock-free mutex, be the anonymous
  registers read/write or read/modify/write.\\~\\

{\bf Keywords}: Anonymous shared memory,
  Asynchronous system,
  Atomic register,
  Complexity, Computability,
  Deadlock-freedom, Mutual exclusion,
  Read/modify/write register,  Read/write register,
  Tight characterization. 
\end{abstract}

\section{Introduction}

\subsection{Memory Anonymity}

While the notion of {\it memory-anonymity} has been implicitly used in
some works in the early eighties (e.g.,~\cite{R82}), it was explicitly
captured as a concept and investigated only very recently
in~\cite{T17}\footnote{See~\cite{RC18} for
an introductory survey to process anonymity and memory anonymity.}.
More precisely, this paper considers the explicit case where ``there
is no a priori agreement between processes on the names of shared
memory locations'', and presents possibility/impossibility
results in such systems.

Considering a shared memory defined as an array $R[1..m]$ of $m$
memory locations (registers), memory-anonymity means
that, while the same location identifier $R[x]$ always denotes the
same memory location for a process $p$, it does not necessarily denote
the same memory location for two different processes $p$ and $q$.
This means that there is an adversary that
may initially give different global names to different processes for the
same memory locations.  In other words, the adversary associates a
permutation on the set of memory indexes $\{1,\cdots,m\}$ with each
process $p$, and $p$ uses them to access the memory locations.  This
is illustrated in Table~\ref{table:memory-anonymous} which considers two
processes $p$ and $q$ and a shared memory made up of three registers
(as an example the register known as $R[2]$ by $p$ and the register
known as $R[3]$ for $q$ are the very same register, which
actually is  $R[1]$ from an external omniscient  observer point of view).

\begin{table}[h!]
\begin{center}
  \begin{tabular}{|c|c|c|}
    \hline
    names for an         & location names    &  location names    \\
    ~external observer~ & ~for process $p$~ &  ~for process $q$~ \\
    \hline
    \hline
    $R[1]$ &  $R[2]$    & $R[3]$  \\
    \hline
    $R[2]$  &  $R[3]$   & $R[1]$  \\
    \hline
    $R[3]$ &  $R[1]$  &  $R[2]$  \\
    \hline
    \hline
    permutation & $2,3,1$ & $3,1,2$\\

    \hline
  \end{tabular}

\end{center}
\caption{An illustration of the anonymous shared memory model}
\label{table:memory-anonymous}
\end{table}

In addition to its possible usefulness in some applications
(e.g.,~\cite{RTB18}) , the memory-anonymous communication model
``enables us to better understand the intrinsic limits for
coordinating the actions of asynchronous processes''~\cite{T17}.

\subsection{Related Work}
This paper originates  from the work described in ~\cite{T17}, where
the memory-anonymous mutex problem is introduced,  and where first
results are presented  on  mutual exclusion
in read/write memory-anonymous systems,  namely
\begin{itemize}
  \item
  a symmetric deadlock-free algorithm for two processes
  (the notion of ``symmetric'' is related to the use of process
  identifiers; it will be defined in Section~\ref{sec:def-symmetric}),
  \item
  a theorem stating there is no  deadlock-free algorithm
  if the number of processes is not known,
  \item
    a necessary condition stating that any symmetric deadlock-free mutex
    algorithm for $n$ processes, which uses $m\geq 2$ read/write
    registers, requires that $m$ belongs to the set
    $M(n)=\{m: \forall~ \ell: 1<\ell \leq n:~ \ggcd(\ell,m)=1\}$ (i.e.,
    $\ell$ and $m$ are relatively prime).  Let us observe that $M(n)$
    is infinite (among other values, it contains an infinite number of
    prime numbers).
\end{itemize}

\subsection{Content of the Paper}
\label{sec:content}
As announced, this paper is on deadlock-free mutual exclusion in an
$n$-process system ($n\geq 2$)  where the processes communicate by
accessing a shared memory composed of $m$ anonymous registers
(hence there are no other  non-anonymous registers).\\

\noindent
{\it Preliminary definitions.}
Two types of registers are considered,
which give rise to two communication models. 
\begin{itemize}
\item
  A read/write (RW) register is an atomic register that
  provides the processes with a read operation and a write
  operation~\cite{L86}.

 In the RW communication model, all the registers are RW registers.
\item
A read/modify/write (RMW) register is an atomic register that
provides the processes with a read operation, a write operation,
and an additional  read/modify/write operation~\cite{HW90}.
Such an operation allows a process to atomically read a register, and,
based on the value read, computes a new value that is assigned to the
register.  One of the most famous  read/modify/write operations
is $\compareandswap()$. When a process invokes
$R.\compareandswap(x,old,new)$, where $R[x]$ is an anonymous register,
it reads the value of the register locally known as $R[x]$, say $v$,
and assigns it the value $new$ if and only if $new=v$.  In this case
it returns $\ttrue$; otherwise, it returns $\ffalse$.

In the  RMW communication  model, all the registers are RMW registers.
\end{itemize}

\noindent
{\it Results presented in the paper.}
\begin{itemize}
\item
RW anonymous communication model.
A deadlock-free mutex algorithm for this
model is presented, which assumes the necessary condition
stated  above, namely, $m\in M(n)$.
As  the condition $m\geq n$
is necessary to solve mutex in a RW non-anonymous system~\cite{BL93},
it remains necessary in an anonymous system.
The very existence  of the proposed
algorithm  shows that the predicate $m\in M(n) \wedge m\geq n$
(which is equivalent to $m\in M(n)\wedge m\neq 1$)
is a tight
characterization of the values of $m$ which allow deadlock-free mutex
algorithm in RW anonymous systems.  In this sense, the proposed
algorithm is space optimal.
\item
RMW anonymous communication model.  A deadlock-free mutex algorithm
for this model is presented, which requires $m\in M(n)$.  As in this
communication model, $m\geq n$ is not a previously known necessary
requirement to solve deadlock-free mutex in a non-anonymous system, we
cannot conclude from the existence of the previous algorithm that
$m\in M(n)$ is a tight characterization of the values of $m$ which
allow deadlock-free mutex algorithm in RMW anonymous systems.

To address this issue, the paper presents a proof that $m\in M(n)$
is actually a necessary and sufficient condition for deadlock-free
mutex in RMW anonymous systems.  (But let us observe that, for the case $m=1$,
an anonymous memory consisting of a single register, is not really anonymous!)
\end{itemize}

Hence, if we eliminate the particular case of an anonymous memory
composed of a single register ($m=1$), deadlock-free mutex for both
the RW model and the RMW model can be solved if and only if
$m\in M(n)$.  The corresponding algorithms differ in their algorithmic
design, and in the fact that, to enter the critical section, the
algorithm for the RW model requires a process to read its identity
from the $m$ anonymous registers, while the algorithm for the RMW
model requires it to read its identity from a majority of the
anonymous registers only.  Hence, while requiring the same
computability assumption (namely $m\in M(n)$ assuming $m\neq 1$),
these algorithms differ from a complexity point of view (measured as
the number of registers that have to contain the same process identity
to allow this process to enter the critical section).

\subsection{Roadmap}

This paper is composed of~\ref{sec:conclusion} sections.
Section~\ref{sec:model-definitions} introduces the computational model
and provides some technical definitions.
Section~\ref{sec:algorithm-rw} presents an $n$-process symmetric
deadlock-free mutex algorithm for the anonymous RW communication
model. Section~\ref{sec:proof-rw} proves its correctness.
Section~\ref{sec:algorithm-rmw} presents an $n$-process
symmetric deadlock-free mutex algorithm for the anonymous RMW
communication model. Section~\ref{sec:proof-rmw-correctness} proves its
correctness,  while Section~\ref{sec:proof-rmw-space}  proves a space
lower bound which shows that the algorithm is space optimal.
Finally, Section~\ref{sec:conclusion} concludes the
paper.  It is important to notice that both algorithms have a first
class noteworthy property, namely, their simplicity.

\section{System Model, Symmetric Algorithm, and Mutual Exclusion}
\label{sec:model-definitions}

\subsection{Processes  and Anonymous Registers}
\label{sec:memory-anonymous}
The system is composed of $n$ asynchronous processes $p_1$, ..., $p_n$
with identifiers from a set $\mathcal{P}$. ``Asynchronous'' means that
each process proceeds in its own speed, which can vary with time and
 always remains unknown to the  processes.

When considering a process $p_i$, $i$ is its index,  which is used only
to distinguish processes from an external point of view.
A process $p_i$ knows its own identity, denoted
$id_i$, but never knows its index $i$.  No two processes have the same identity.

Each process knows the number, $n$, of processes in the system, and
all processes know a common symbol $\bot\notin\mathcal{P}$, which is
interpreted as a default identity (hence, when it reads an anonymous register,
a process can distinguish a process identity  from~$\bot$).

~\\
\noindent
RW {\it anonymous communication model.}\\
Processes communicate through a memory anonymous array $R[1..m]$,
which can be accessed by two operations, denoted $R.\wwrite{}$ and
$R.\rread{}$.
As already indicated, memory-anonymity means that, for each process
$p_i$, there is a permutation $f_i()$ over the set
$\{1,\dots,m\}$ such that, when $p_i$ uses address $R[x]$, it actually
accesses $R[f_i(x)]$.  Anonymity means that no permutation $f_i$ is
known by any process; each $f_i$ is defined by an adversary.

\begin{itemize}
  \item
  When a process $p$ invokes $R.\wwrite(x,v)$ it writes value
  $v$ in the atomic read/write register $R[f_i(x)]$,
  \item
  When a process $p$ invokes $R.\rread(x)$ it obtains the value
  currently saved in the register locally denoted $R[f_i(x)]$.
\end{itemize}

To simplify the presentation of the first  algorithm, we also assume that
process $p_i$ can use an operation $R.\ssnapshot()$ to obtain the value of
array $\big[R[f_i(1)],\dots,[R[f_i(m)]\big]$ as if the read of all its
  entries where instantaneous (i.e., produced at a single point of the
  time line during the operation~\cite{HW90,L86}).  We require that
  the operation $\ssnapshot()$
  satisfies the following progress guarantee:
\begin{equation}\label{eq:scan-progress}
\parbox{\dimexpr\linewidth-6em}{ In any interval of the execution
  throughout which no process calls $\wwrite()$, any invocation of
  $R.\ssnapshot()$ by a process $p_i$ terminates within a finite
  number of $p_i$'s steps~\cite{H91}.}
\end{equation}

The memory-anonymous $\ssnapshot()$ operation  is a simple extension of the
classical $\ssnapshot()$ introduced in~\cite{AADGMS93,A94}.  All its
executions are linearizable~\cite{HW90}.  The operation $R.\ssnapshot()$
can be implemented with the well-known ``double scan'' technique (as
used in~\cite{AADGMS93}), where each process $p_i$ is provided with an
additional local sequence number $sn_i$, which it uses to identify all
its write invocations (namely, when it invokes $R.\wwrite(x,v)$, $p_i$
actually issues the following sequence of statements
``$sn_i \leftarrow sn_i+1$; $R[x] \leftarrow (v,id_i,sn_i)$''.  As no two
processes have the same identity, each invocation of $R.\wwrite()$ is
unambiguously identified.\footnote{\label{footnote-label} The proof of
  the operations $R.\wwrite()$, $R.\rread()$, and $R.\ssnapshot()$
  terminate and are linearizable is the same as the one done
  in~\cite{AADGMS93}.  As far as the proof of $R.\ssnapshot()$ is
  concerned, this comes from the observation that, in the algorithm
  described in~\cite{AADGMS93}, the order in which a process scans the
  array $R[1..m]$ is irrelevant.  The important point is that, after
  it sequentially scanned twice the array $R[1..m]$, a process
  compares the corresponding entries of the two copies of $R[1..m]$ it
  has obtained.  In our case, for any $x$, the first reading of $R[x]$
  and a second reading of $R[x]$ by the same process are on the very
  same memory location, from which follows that it correctly compares
  the corresponding entries of the two copies it has obtained to see
  if $R[1..m]$ changed between the two consecutive scans.}  To not
overload the presentation, the sequence numbers associated with the
write operations (and used in the ``double scan'' inside the snapshot
operations) are left implicit in the rest of the paper\footnote{Defining
  each register as a record which has two fields (a value field and a
  sequence number field)  with global (non-anonymous) names is done only
  for convenience. The two values in these fields can be encoded as
  a single value, removing the need for using more than one field.}.

~\\
\noindent
RMW {\it anonymous communication model.}\\
This is the RW model where, in addition to  $R.\rread(x)$ $R.\wwrite(x,v)$,
a process can also invoke the operation  $R.\compareandswap$($x,-,-$)
defined in Section~\ref{sec:content}.
(The deadlock-free mutex algorithm described in Section~\ref{sec:algorithm-rmw}
does not use the operation  $R.\ssnapshot()$.)

\subsection{Symmetric Algorithm}
\label{sec:def-symmetric}
The notion of a {\it symmetric algorithm} dates back to the
eighties~\cite{GG90,JS85}. Here, as in~\cite{T17}, a {\it symmetric
  algorithm} is an ``algorithm in which the processes are executing
exactly the same code and the only way for distinguishing processes is
by comparing identifiers. Identifiers can be written, read, and
compared, but there is no way of looking inside an identifier. Thus it
is not possible to know if whether an identifier is odd or even''.

Two variants of symmetric algorithms can be considered,
\emph{symmetric with equality} and \emph{symmetric with arbitrary
  comparisons}.  As in~\cite{T17} we only consider the more
restricted version, symmetric with equality, where the only comparison
that can be applied to identifiers is equality.  In particular, there
is no order structuring on  the identifier name space.  Throughout this
paper, symmetric refers to symmetric with equality.

Moreover, in order for the initial values not to be used to destroy anonymity
(which could favor a given process),  all registers are initialized
to the same value, namely the default value $\bot$.

\subsection{Mutual Exclusion}

Mutual exclusion is the oldest (and one of the most important)
synchronization problem.  Formalized by Dijkstra in the
mid-sixties~\cite{D65}, it consists in building what is called a lock
(or mutex) object, defined by two operations, denoted $\llock()$ and
$\uunlock()$.  (Recent textbooks including mutual exclusion and
variants of it are~\cite{R13,T06}.)

The invocation of these operations by a process $p_i$
always follows the following pattern: ``$\llock()$; {\it critical
  section}; $\uunlock()$'', where ``critical section'' is any sequence of
code.  A process that is not in the critical section and has no
pending $\llock()$ or $\uunlock()$  invocation, is said to be in the
\emph{remainder section}.  An infinite execution is \emph{fair}, if
every process that has a pending $\llock()$  or $\uunlock()$
invocation, either
finishes its operation or executes infinitely many steps.  The mutex
object satisfying the deadlock-freedom progress condition is defined
by the following two properties.
\begin{itemize}
  \item Mutual exclusion: No two processes are simultaneously in their
    critical section.
  \item Deadlock-freedom: If a process $p_i$ has a pending $\llock()$ or
    $\uunlock()$ invocation and no process is in the critical section, then
    some process $p_j$ (possibly $p_j\neq p_i$) eventually finishes its
    $\llock()$ or $\uunlock()$ operation, provided the execution is fair.
\end{itemize}

\section{RW Anonymous Model: Symmetric Deadlock-Free Mutex}
\label{sec:algorithm-rw}
This section presents Algorithm~\ref{algo:mutex-memory-anonymous-algorithm-rw},
which is a symmetric (with respect to equality) deadlock-free mutex algorithm
suited to the RW memory-anonymous communication model.
As indicated in the introduction, as this  algorithm works for
the necessary condition $(m>1)\wedge(m\in M(n))$, its existence proves
that this condition is also sufficient.

\subsection{Representation of the Lock Object}

\textbf{Shared memory:}\quad
Let $m$ be  such that $m>1$ and $m\in M(n)$.
The shared memory is composed of a memory-anonymous array
$R[1..m]$, as defined in Section~\ref{sec:memory-anonymous}.

For any $x\in\{1,\dots,m\}$, the initial value of a register $R[x]$ is
$\bot$.  If $R[x]\neq\bot$, it contains the identity of the last
process that wrote in this register.  From a terminology point of
view, we say
\begin{itemize}
\item ``process $p_i$ owns $R[x]$'', if $R[x] = id_i$;
\item ``register $R[x]$ is owned'' if $R[x]\neq\bot$;
\item ``$R$ is full'' if all its entries are owned; and
\item ``$R$ is empty'' if none of its entries are owned.
\end{itemize}

\noindent\textbf{Local memory:}\quad Each process $p_i$ manages two
local variables: an integer  denoted  $cnt_i$, and a
local array denoted  $view_i[1..m]$.  The aim of $view_i[1..m]$ is
to contain the value of $R$ obtained by $p_i$ from its last invocation
of $R.\ssnapshot()$.  To prevent confusion, the shared array $R[1..m]$
is denoted with an uppercase letter, while the local variables are
denoted with lowercase letters.  As already indicated, the local
sequence number associated with each write operation of a process $p_i$
is left implicit.

\subsection{Algorithm}

\subsubsection{Underlying Principles}
The core of Algorithm~\ref{algo:mutex-memory-anonymous-algorithm-rw}
consists in managing a competition
among the processes until all the entries of $R[1..m]$ contain the
same process identity, the corresponding process being the winner.

When a process invokes $\uunlock()$, or when it concludes while
competing that it will not be the winner, it resets the entries of
$R[1..m]$ containing its identity to $\bot$ (their initial value).

Hence, the core of the algorithm  lies in the definition of
predicates that direct a process to either withdraw from the
competition or continue competing.  To this end, a process $p_i$
checks whether its local view
$view_i$  (obtained from the invocation of
$R.\ssnapshot()$) is full (line~\ref{rw-05}), and if so, whether $p_i$ owns less than
the average of all registers present in the competition (line~\ref{rw-09}).

\begin{algorithm}[ht]
\centering{
\fbox{
\begin{minipage}[t]{150mm}
\footnotesize
\renewcommand{\baselinestretch}{2.5}
\resetline
\begin{tabbing}
aaaaa\=aa\=aaa\=aaa\=aaaaa\=aaaaa\=aaaaaaaaaaaaaa\=aaaaa\=\kill 

 \>   $m>1$  and
   $ m\in \{m \mbox{ such that } \forall ~\ell \in \{2,...,n\}:\ggcd(\ell,m)=1\}$\\
 \>   $R[1..m]$: array of anonymous RW atomic registers, each initialized to $\bot$\\
 \>   $p_i$:  process executing this code; $id_i$ is its identity\\
 \>   $view_i$: process $p_i$'s local array of size $m$ (with global scope)\\

 -------------------------------------------------------------------------------------------------------------------------------------\\
 {\bf operation}
 $\oowned()$ {\bf is}\\
 \line{rw-01}  \>
 $\rreturn~(|\{ x\in \{1,\dots,m\}:view_i[x]=id_i\}|)$.
 \% $\#$ of registers owned by $p_i$  \%\\

 -------------------------------------------------------------------------------------------------------------------------------------\\

 {\bf operation}  $\sshrink()$ {\bf is}\\
 \line{rw-02}  \>
      {\bf for each} $x$ {\bf such that}  $view_i[x]=id_i$ {\bf do}

{\bf if} $(R.\rread(x)=id_i)$ {\bf then} $R.\wwrite(x,\bot)$ {\bf end if}
{\bf end for}.\\%~\\

 -------------------------------------------------------------------------------------------------------------------------------------\\

 {\bf operation}  $\llock()$ {\bf is}\\
 \line{rw-03}  \>         {\bf repeat}\\

\line{rw-04} \>\>        {\bf repeat}  $view_i \leftarrow R.\ssnapshot()$
     {\bf until} $\oowned() >0 \vee \forall~ x\in \{1,\dots,m\}: view_i[x]=\bot$ {\bf end repeat};\\

\>\>  \%  This point is reached only if either $p_i$ is present (at least one entry of $R$
     contains $id_i$) or no one is \%\\

\line{rw-05}  \>\>     {\bf if}  ($\exists~ x\in \{1,\dots,m\}: view_i[x]=\bot$)\\

\line{rw-06}  \>\>\> {\bf then}~ \=$R.{\sf write}(x,id_i)$\\

       \line{rw-07}  \>\> \>   {\bf else} \>
\% $view_i$ is full \%\\

\line{rw-08}  \>\>\>\>
let $cnt_i = |\{view_i[1], \dots,view_i[m]\}|$; \% number of current competitors \%\\

\line{rw-09}  \>\>\>\>
     {\bf if} \=  ($\oowned() < m / cnt_i$) {\bf then} $\sshrink()$  {\bf end if}\\

 \>\>\>\>\>
\% $p_i$ owns fewer registers than the average   $\Rightarrow$ $p_i$ withdraws from the competition \%\\

\line{rw-10} \> \>  {\bf end if}\\

\line{rw-11} \> {\bf until} $\forall~ x\in \{1,\dots,m\}: view_i[x]=id_i$ {\bf end repeat}.\\

 -------------------------------------------------------------------------------------------------------------------------------------\\

{\bf operation}  $\uunlock()$ {\bf is}\\
\line{rw-12} \>  $\sshrink()$.

\end{tabbing}
\normalsize
\end{minipage}
}
\caption{Algorithm 1: RW memory-anonymous deadlock-free  mutex
                  ($n$-process system, $n\geq 2$, code for $p_i$)}
\label{algo:mutex-memory-anonymous-algorithm-rw}
}
\end{algorithm}

\subsubsection{Detailed Algorithm Description}

Operation $\uunlock()$ is a simple invocation of an internal operation
called $\sshrink()$ (line~\ref{rw-12}).  With a $\sshrink()$ invocation,
a process $p_i$ removes itself from the array by considering its latest
view, $view_i$.  More specifically, for each index $x\in\{1,\dots  m\}$
with $view_i[x]=id_i$, process $p_i$ first reads $R[x]$, and if $R[x]$
still equals $id_i$,  it writes  $\bot$ into $R[x]$
(line~\ref{rw-02}).

The core of the algorithm is the code of the operation $\llock()$.  When
a process $p_i$ invokes this operation, it enters a ``repeat-until''
loop from which it can only exit when it obtains a snapshot of the
anonymous shared memory $R[1..m]$, according to which $p_i$ owns all
entries (lines~\ref{rw-03}-\ref{rw-11}).

Hence, process $p_i$ first repeatedly invokes $R.\ssnapshot()$
(line~\ref{rw-04}) until it sees only $\bot$ in the array $view_i$
obtained from $R.\ssnapshot()$ (which means that, from its local point
of view, there is no competition), or it is present in this array
(which means it is already competing).  Then, when it stops looping,
$p_i$ checks whether $view_i$ is full (line~\ref{rw-05}) to know if it
should continue writing (line~\ref{rw-06}) or if it should consider
withdrawing from the competition (lines~\ref{rw-07}-\ref{rw-09}).

If $view_i$ is full, processes are
engaged in a competition.  If its identity appears in fewer than the
average number of owned registers, process $p_i$ withdraws from the
competition by invoking the operation $\sshrink()$
(lines~\ref{rw-07}-\ref{rw-09}), which
suppresses its identity from the anonymous RW  memory $R[1..m]$.  After
finishing its $\sshrink()$  invocation, a process re-enters the repeat-until
loop at line~\ref{rw-04}.  The fact that $m\in M(n)$
guarantees that not all processes that appear in $R$ when it is
full, own the same number of registers, so at least one process will
withdraw.  If a process owns at least the average number of registers
when its view is full,  it re-enters the
repeat-until loop and invokes  the operation
$\ssnapshot()$ again at line~\ref{rw-04}.

If $view_i$ is not full and $p_i$ owns at least one register, it continues
competing.  To this end, before re-entering the repeat-until loop,
$p_i$ chooses an entry of $R[1..m]$ equal to $\bot$, and writes its
identifier $id_i$  in this register (lines~\ref{rw-05}-\ref{rw-06}).\\

To summarize, during a $\llock()$ operation, a process $p_i$ decides its
future steps based on its latest view of the anonymous memory  as follows:
\begin{enumerate}
  \item If $p_i$ owns all registers,  it enters the critical section
    (line~\ref{rw-11}).
  \item If $p_i$ owns no register, and the view is not empty, then it
    waits (by repeatedly taking snapshots) until it obtained an empty
    view (line~\ref{rw-04}).
  \item If the view is full, and $cnt_i$ different processes own some
    registers, and $p_i$ owns fewer than $m/cnt_i$ registers, then it
    removes itself from all registers it owns by calling $\sshrink()$
    (line~\ref{rw-09}).
  \item If the view is not full,
    there is at least one register that is not owned, and $p_i$ writes its
    identity $id_i$ into any not owned  register (line~\ref{rw-06}).
\end{enumerate}

\section{RW  Model:
  Proof of Algorithm~\ref{algo:mutex-memory-anonymous-algorithm-rw}
  and  Tight  Space  Bound}
\label{sec:proof-rw}

\subsection{Proof of Algorithm~\ref{algo:mutex-memory-anonymous-algorithm-rw}}

Let us remember that the anonymous RW array $R[1..m]$ is the only
object that the processes can use to communicate.  The notions of
``time'', ``first'' and  ``last'' used in the proofs are well-defined, as all
$\wwrite()$ and $\rread()$ operations are atomic.  As stated in
Section~\ref{sec:memory-anonymous}, the proof assumes that operation
$\ssnapshot()$ (which can be implemented from atomic read/write operations)
is linearizable and satisfies the progress condition
\ref{eq:scan-progress}.
 The proof assumes $n\geq 2$, as otherwise mutual exclusion is trivial.
Moreover, let us remember that  $m$ is assumed to be be greater than $1$ and
belong to the set
$M(n)=\{m: \forall~ \ell: 1<\ell \leq n:~ \ggcd(\ell,m)=1\}$, from which
follows that $m>n$.

Let $E$ be  an arbitrary infinite execution $E$,  $L(E)$ 
an execution where all $\ssnapshot()$operations occur atomically at
their linearization points (i.e., $L(E)$ is a linearization of all
operations on $R$ in $E$).

\begin{theorem}
  \label{theo:mutex}
  Algorithm~{\em{\ref{algo:mutex-memory-anonymous-algorithm-rw}}}
  satisfies mutual exclusion.
\end{theorem}
\begin{proofT}
  Consider history $L(E)$.  Let us suppose by contradiction that two
  processes are inside their critical section at the same time, and
  assume that $p_i$ is the first of them to take its last snapshot
  before entering its critical section. More precisely,
  suppose process $p_i$'s $\llock()$ invocation
  terminates (and thus $p_i$ enters the critical section) following some
  iteration of the outer repeat-until loop in $\llock()$.  Then due to
  the predicate of line~\ref{rw-11}, $p_i$ owns all registers of
  $R$ at the point of $p_i$'s $\ssnapshot()$  (line~\ref{rw-04}) in its
  last iteration.  Therefore, in the same iteration the predicate of
  line~\ref{rw-05} and  the predicate of
  lines~\ref{rw-09} are false, and the  predicate of
  lines~\ref{rw-11} is true.
   Hence, the $\ssnapshot()$  in line~\ref{rw-04} at
  the beginning of the iteration is $p_i$'s last access to the RW
  anonymous memory before its $\llock()$ operation terminates.  We
  therefore say a {\it process enters the critical section} at the point
  when it is taking a snapshot in line~\ref{rw-04} while owning all
  registers.

  Now suppose $p_i$ enters the critical section at some point $t$.  Let
  $t'$ be the point when $p_i$ executes its first shared memory
  operation in its subsequent $\uunlock()$ invocation, if there is such
  an invocation, and otherwise $t'=\infty$.  We prove below the following
  claim:
  \begin{claim}
   \label{claim:mutex}
    \text{Throughout $[t,t']$, all invocations of $\ssnapshot()$ contain
    the identity of $p_i$.}
  \end{claim}
 It follows from this claim that at no point in $[t,t']$ a process
 other than $p_i$ can observe itself as owning all registers.  Also,
 as assumed at the beginning of the proof, process $p_i$ is the
 \emph{first} to take its last snapshot before entering its critical
 section.  Thus no other process, except $p_i$, can be in the critical
 section throughout $[t,t']$, which contradicts the assumption that
 $p_i$ is not alone in the critical section.\\

 Proof of the claim.
 For the purpose of a contradiction, let us
 assume that Claim~\ref{claim:mutex}  is not true.
Because all $m$ registers are owned by $p_i$ at time $t$ and $m>n$,
by the pigeonhole principle, at least one process
has issued more than one write that changed the value of a register
from the identity of $p_i$ to another value.
Let $p_j$ be the first process to do so.
 Hence,  process $p_j$ took a snapshot at some point $T_s\in [t,t']$ in
  line~\ref{rw-04} at the beginning of the iteration of the outer
  repeat-until loop in which it executes its second $R.\wwrite()$ in
  line~\ref{rw-05}, that changes the value of a register
from the identity of $p_i$ to another value.

  Process $p_i$ is the only process that can write its own identity,
  it owns all the registers at time $t$, and it does not execute any
  write operation in $[t,t']$.
  Then, in the snapshot taken at  $T_s\in [t,t']$ by $p_j$,
  the second register overwritten by $p_j$ contains $p_i$'s identity,
  and is not chosen at line~\ref{rw-05}.
A contradiction which completes the proof of the claim and the theorem.
\renewcommand{\toto}{theo:mutex}
\end{proofT}

\begin{theorem}\label{theo:deadlock-freedom}
  Algorithm~{\em{\ref{algo:mutex-memory-anonymous-algorithm-rw}}}
  is deadlock-free.
\end{theorem}

The remainder of this section is devoted to the proof of this theorem.
For the purpose of contradiction let us  assume that $E$ is an infinite fair
execution, and that after some point $t^\ast$ no
invocation of $\llock()$ or $\uunlock()$
 terminate, even though at least one process has a pending
 $\llock()$ or $\uunlock()$  operation  and no process is in the critical
section.  Since $\uunlock()$ is wait-free~\cite{H91} (see also
the Claim ~\ref{clm:shrink-works} below) we may assume w.l.o.g. that no
invocation of $\uunlock()$ is pending at any point after $t^\ast$.

\Xomit{
~\\
\noindent
{\it Definitions}.
\begin{itemize}
  \item A process is \emph{shrinking} if it is poised to read or write in a
  $\sshrink()$ operation.
  \item A process is  \emph{large} if it owns at least $2$ registers.
\end{itemize}
}

\begin{claim}\label{clm:shrink-works}
  In any execution, each invocation of $\sshrink()$ by a process $p_i$
  terminates within a finite number of $p_i$'s steps, and when it
  does, process $p_i$ owns no register.
\end{claim}
\begin{proofCL}
  Process $p_i$ executes at most $m$ iterations of the for-loop in
  $\sshrink()$, and in each iteration it executes the wait-free
  operations $\wwrite()$  and $\rread()$, so $\sshrink()$ terminates after a
  finite number of $p_i$'s steps.  Before calling $\sshrink()$, process $p_i$
  calls $\ssnapshot()$ to obtain $view_i$ (line~\ref{rw-06}), and
  it does not write to the RW anonymous memory $R[1..m]$ between
  that  $\ssnapshot()$ and its subsequent  call of $\sshrink()$.
  In the for-loop in the operation   $\sshrink()$,  process $p_i$
  writes $\bot$ into all registers $R[x]$, $x\in\{1,\dots,m\}$,
  such that $view_i[x]=id_i$.  Since no other process writes the value  $id_i$
  to  any register $R[x]$, no register contains $id_i$ anymore when $p_i$
  terminates its  $\sshrink()$ operation.
 \renewcommand{\toto}{clm:shrink-works}
\end{proofCL}

\begin{claim}\label{clm:remainder-section-owns-no-register}
  If a process owns a register, then it is not in the remainder
  section (i.e., it has a pending $\llock()$ or $\uunlock()$
  invocation, or it is  in the critical section).
\end{claim}
\begin{proofCL}
   A process $p_i$ can only begin to own a register $R[x]$,
   $x\in\{1,\dots,m\}$, by writing $id_i$ into $R[x]$, that can only
   happen in line~\ref{rw-12} of the operation $\llock()$.  When
   $p_i$ enters the remainder section, it has not written to any
   register of $R$ since its latest $\sshrink()$ invocation in the
   operation $\uunlock()$ terminated, so the statement follows from
   the Claim~\ref{clm:shrink-works}.
\renewcommand{\toto}{clm:remainder-section-owns-no-register}
\end{proofCL}

In the following, for any given time $t\geq t^\ast$, we say that
$p_i$ is {\it competing} if $p_i$ has a pending $\llock()$ operation at $t$
and  the last snapshot taken by $p_i$ before $t$ satisfies the condition at
line~\ref{rw-04} (i.e. $p_i$ is not stuck in the inner loop).

\begin{claim}\label{clm:no-shrink}
 At any point $t\geq t^\ast$, there is a competing process $p_i$ whose
 last $\ssnapshot()$ invocation does not cause it to invoke
 $\sshrink()$ at line~\ref{rw-09}.
\end{claim}
\begin{proofCL}
If at least one competing process $p_i$ obtains a view that is not
full, the condition at line~\ref{rw-05} is satisfied, and thus this
view does not cause $p_i$ to invoke $\sshrink()$. We can then consider
that all competing processes have obtained a full view in their last
snapshot.

Let $p_i$ be the process that owns the most registers in the last
snapshot taken before $t$ (if more than one process satisfy this
condition, $p_i$ can be chosen arbitrarily among them).  If $p_i$ took
this snapshot, it wouldn't cause it to invoke $\sshrink()$ ($p_i$ owns
more than the average, condition at line~\ref{rw-09}).  Let us then
consider that $p_i$ didn't take this last snapshot before $t$, but
took one previously at time $t'<t$.  Process $p_i$ is the only one
which can write its own identity, and its last view was full, causing
it not to write (condition at line~\ref{rw-04}).  At time $t'$, $p_i$
then owns at least as many registers as in the last snapshot taken
before $t$. Furthermore, any competing process in the last snapshot
taken before $t$ is also competing at time $t'$ (otherwise it would be
stuck in the inner loop at line~\ref{rw-04}).  Thus, the view taken by
$p_i$ at time $t'$ does not satisfy the condition at line~\ref{rw-09},
and does not cause $p_i$ to invoke $\sshrink()$.
\renewcommand{\toto}{clm:no-shrink}
\end{proofCL}

\begin{claim}\label{clm:has-to-shrink}
  At any point $t\geq t^\ast$, if there is more than one competing
 process, at least one of them will invoke $\sshrink()$.
\end{claim}
\begin{proofCL}
Suppose not. Note that the only point at which a process can write
$\bot$ is during the $\sshrink()$ operation.  If at least one
competing process obtains a view that is not full, it will invoke
$R.\wwrite()$.  This will happen again until no register has the value
$\bot$ and all competing processes obtain full views in their last
snapshot, preventing them from writing.  We can then consider
w.l.o.g. that, at time $t$, all competing processes have stopped
writing and have obtained the same view.  Let $cnt$ be the number of
these competing processes.

Because $1 < cnt \leq n$ and $\forall~ \ell: 1<\ell \leq n:~
\ggcd(\ell,m)=1$, at least one competing process owns less registers
than $m/cnt$, causing it to call $\sshrink()$; a contradiction which
proves the claim.
\renewcommand{\toto}{clm:has-to-shrink}
\end{proofCL}

\begin{proofT} of Theorem~\ref{theo:deadlock-freedom}.

By Claim~\ref{clm:no-shrink}, at any point $t\geq t^\ast$, there is a
competing process whose last $\ssnapshot()$ invocation does not cause
it to invoke $\sshrink()$.  By Claim~\ref{clm:shrink-works}, any
$\sshrink()$ operation terminates, and causes the invoking process to
be stuck in the inner loop at line~\ref{rw-04}, causing it to stop
competing after its next $\ssnapshot()$ invocation.  This implies that
at least one competing process never calls $\sshrink()$ after point $t^\ast$.

By assumption, no process is in the critical section, and no
$\uunlock()$ operation is pending.  By
Claim~\ref{clm:remainder-section-owns-no-register}, if a process owns
a register, then it is not in the remainder section. The only
processes that own registers are then the ones that are competing.

By Claim~\ref{clm:has-to-shrink}, if there is more than one competing
process, at least one of them invokes $\sshrink()$, causing it to stop
competing.  There is then eventually a single competing process that
owns all the registers, a contradiction with the original assumption
that after some point $t^\ast$, no invocation of $\llock()$ or
$\uunlock()$ terminates, even though at least one process has a
pending $\llock()$ or $\uunlock()$ operation and no process is in the
critical section.
\renewcommand{\toto}{theo:deadlock-freedom}
\end{proofT}

\subsection{RW Memory-Anonymous Model:  Tight   Space   Bound}
Given  $M(n)=\{m: \forall~ \ell: 1<\ell \leq n:~ \ggcd(\ell,m)=1\}$,
it is shown in~\cite{T17} that $m\in M(n)$ is a necessary
condition for any algorithm solving symmetric deadlock-free mutex
in an anonymous memory composed of $m$ read/write registers.
As already indicated, as  $m\geq n$ is a  necessary
condition for any algorithm solving deadlock-free mutex
in a non-anonymous system, it remains necessary in a read/write
anonymous system.  This translates as follows:
$m\in M(n)\setminus \{1\}$  is a necessary condition
for  deadlock-free mutex
in an anonymous memory composed of $m$ read/write registers.

As Algorithm~\ref{algo:mutex-memory-anonymous-algorithm-rw}
solves   deadlock-free mutex under this condition, it follows that
$m\in M(n)\setminus \{1\}$ is a necessary and sufficient condition.

\section{RMW Anonymous  Model: Symmetric Deadlock-Free Mutex}
\label{sec:algorithm-rmw}

This section presents an algorithm that implements a deadlock-free
mutex lock object in an $n$-process RMW memory-anonymous system.
As the previous algorithm, the algorithm presented below
is particularly simple.

\subsection{Representation of the Lock Object}
The shared anonymous memory is made up of $m$ RMW atomic registers,
denoted $R[1..m]$ where $m\in
\{1\} \cup \{m: \forall~ \ell: 1<\ell \leq n:~ \ggcd(\ell,m)=1\}$
(let us notice that  this set includes the value $1$).

In Algorithm~\ref{algo:mutex-memory-anonymous-algorithm-rmw}, a process
uses three local variables, denoted $most\_present_i$, $owned_i$,
and $view_i$ (which has the same meaning as in
Algorithm~\ref{algo:mutex-memory-anonymous-algorithm-rw}).

\begin{algorithm}[ht]
\centering{
\fbox{
\begin{minipage}[t]{150mm}
\footnotesize
\renewcommand{\baselinestretch}{2.5}
\resetline
\begin{tabbing}
aaaaa\=aa\=aaa\=aaa\=aaaaa\=aaaaa\=aaaaaaaaaaaaaa\=aaaaa\=\kill 

 \>
$m\in \{m \mbox{ such that } \forall ~\ell \in \{2,...,n\}:\ggcd(\ell,m)=1\}$\\
 \>   $R[1..m]$: array of anonymous RMW atomic registers,
     each initialized to $\bot$\\
 \>   $p_i$:  process executing this code; $id_i$ is its identity\\
 \>   $view_i$: process $p_i$'s local array of size $m$ (with global scope)\\
 -----------------------------------------------------------------------------------------------------------------------------\\

{\bf operation}
 $\oowned()$ {\bf is}
 $\rreturn~(|\{ x\in \{1,\dots,m\}:view_i[x]=id_i\}|)$.
      \% $\#$ of registers owned by $p_i$  \%\\

 -----------------------------------------------------------------------------------------------------------------------------\\
{\bf operation}  $\llock()$ {\bf is}\\
 \line{rmw-01}  \>  {\bf repeat}\\

 \line{rmw-02} \>\>
 {\bf for each} $x\in \{1,...,m\}$
 {\bf do} $R.\compareandswap(x,\bot,id_i)$ {\bf end for};\\

 \line{rmw-03} \>\>
{\bf for each} $x\in \{1,...,m\}$
 {\bf do}  $view_i[x]  \leftarrow R.\rread(x)$   {\bf end for};\\

 \line{rmw-04} \>\>
 $most\_present_i \leftarrow$
      maximum number of times the same non-$\bot$ value appears in $view_i$;\\

\line{rmw-05} \>\>
      $owned_i \leftarrow \oowned()$;\\

\line{rmw-06} \>\>
{\bf if} $owned_i < most\_present_i$ {\bf then}\\

\line{rmw-07} \>\>\> {\bf for each} $x\in \{1,...,m\}$  {\bf do}
 {\bf if} $(view_i[x]= id_i)$  {\bf then} $R.\wwrite(x,\bot)$   {\bf end if}
 {\bf end for};\\

\line{rmw-08} \>\>\> {\bf repeat}\\

\line{rmw-09} \>\>\>\> {\bf for each}
 $x\in \{1,...,m\}$
 {\bf do}  $view_i[x]  \leftarrow R.\rread(x)$   {\bf end for}\\

 \line{rmw-10} \>\>\>
      {\bf until} $\forall~ x\in \{1,\dots,m\}: view_i[x]=\bot$ {\bf end repeat}\\

\line{rmw-11} \> \>  {\bf end if}\\
\line{rmw-12} \> {\bf until} $owned_i>m/2$ {\bf end repeat}.\\

 -----------------------------------------------------------------------------------------------------------------------------\\
{\bf operation}  $\uunlock()$ {\bf is}\\
\line{rmw-13} \>
{\bf for each} $x\in \{1,...,m\}$
 {\bf do} $R.\compareandswap(x,id_i,\bot)$ {\bf end for}.
\end{tabbing}
\normalsize
\end{minipage}
}
\caption{Algorithm 2: RMW memory-anonymous deadlock-free  mutex
                  ($n$-process system, $n\geq 2$, code for $p_i$)}
\label{algo:mutex-memory-anonymous-algorithm-rmw}
}
\end{algorithm}

\subsection{Algorithm}

When a process $p_i$ invokes $\llock()$, it enters repeat loop that it will
exit when it will obtain a view $view_i[1..m]$
in which its own identity appears in a majority of anonymous registers
(line~\ref{rmw-12}).

Process $p_i$ first invokes the read/modify/write operation
$\compareandswap()$ on all registers in order to write its identity in
all the registers whose current value is the default value $\bot$
(line~\ref{rmw-02}).  Then, it reads (asynchronously) the whole
anonymous memory and saves it in $view_i[1..m]$ (line~\ref{rmw-03}).
From this non-atomic view of the shared memory, $p_i$ computes the
occurrence number of the most present value ($most\_present_i$,
line~\ref{rmw-04}) and the occurrence number of its own identity
($owned_i$, line~\ref{rmw-05}).
\begin{itemize}
\item
  If $owned_i\geq most\_present_i$, $p_i$ proceeds to the next iteration
  of the repeat-until loop  if $owned_i\leq m/2$,
  and enters the critical section if  $owned_i > m/2$.
\item If $owned_i < most\_present_i$, $p_i$ resigns from the competition.
  To this end,  it first writes $\bot$ in all entries that,
  from its local point  of view, contain its identity (line~\ref{rmw-07}),
  and then waits until it sees that all the anonymous registers contain
  the default value $\bot$ (lines~\ref{rmw-08}-\ref{rmw-10}).
\end{itemize}

When a process $p_i$  invokes $\uunlock()$, it simply resets to $\bot$
all the registers that contain its identity $id_i$ (line~\ref{rmw-13}).

\section{RMW Model: Proof of
  Algorithm~\ref{algo:mutex-memory-anonymous-algorithm-rmw}
and Tight  Space Lower  Bound}
\label{sec:proof-rmw}

\subsection{Proof of
  Algorithm~\ref{algo:mutex-memory-anonymous-algorithm-rmw}}
\label{sec:proof-rmw-correctness}

\begin{theorem}\label{theo-algo-2-mutex}
  Algorithm~{\em\ref{algo:mutex-memory-anonymous-algorithm-rmw}}
  satisfies mutual exclusion.
\end{theorem}
\begin{proofT}
Assume that a process $p_i$ is in its critical section,
while some other process,  say process $p_j$, is
executing the operation $\llock()$.
Before $p_i$ entered its critical section the
exit predicate of line~\ref{rmw-12}, namely,
$owned_i > m/2$ must be evaluated to true.
This means that, before $p_i$ entered its critical section,
it succeeded to change more than $m/2$ RMW anonymous registers
from $\bot$  to its identifier $id_i$.
As long as process $p_i$ does not set these RMW registers back to $\bot$,
process $p_j$ cannot succeed in changing more than $m/2$ registers
from $\bot$  to $id_j$.
Thus, process $p_j$ will not be able to enter its critical section while
$p_i$ is in its critical section. 
\renewcommand{\toto}{theo-algo-2-mutex}
\end{proofT}

\begin{theorem}
  \label{algo-2-deadlock-free}
  Algorithm~{\em\ref{algo:mutex-memory-anonymous-algorithm-rmw}}
  is deadlock-free.
\end{theorem}

\begin{proofT}
We show that if a process is trying to enter its
critical section, then some process eventually enters its critical section.

In the first for loop (line~\ref{rmw-02})
each process scans the $m$ RMW anonymous registers trying to set those
that are $\bot$ to its identifier.
If the process is running alone, it will clearly succeed to set
them all to its identifier and will enter its critical section.

When there is contention (i.e., several processes are in their entry
codes) since $\forall ~x\in\{1,...,n\}:$ $m$ and $x$ are relatively
prime, at least one of the processes $p_k$ must find that less than
$most\_present_k$ of the RMW registers are set to its identifier.  It
follows from lines~\ref{rmw-06}-\ref{rmw-07} that $p_k$ gives up the
competition, and waits in the inner repeat loop
(lines~\ref{rmw-08}-\ref{rmw-10}). This enables at least one other
process $p_j$, for which $most\_present_j$ of the RMW registers are
set to its identifier, to proceed.  Repeating this argument,
eventually one of the processes will find that its identifier appears
in more than $m/2$ of the RMW registers and will enter its critical
section.

Finally,  as in its exit code  (line~\ref{rmw-13}),
a process sets to $\bot$ all the registers containing  its identifier.
This enables a possibly waiting process to continue.
Thus, it is not possible for all the processes to simultaneously
remain forever in the operation $\llock()$.
\renewcommand{\toto}{algo-2-deadlock-free}
\end{proofT}

\subsection{RMW Anonymous  Model: Tight  Space Lower Bound}
\label{sec:proof-rmw-space}
\begin{theorem}\label{theo-bound-RMW}
There is an $n$-process symmetric deadlock-free mutual exclusion
algorithm using $m\geq 1$ anonymous RMW registers only if
 $m\in M(n)=\{m: \forall~ \ell: 1<\ell \leq n:~\ggcd(\ell,m)=1\}$.
\end{theorem}

\begin{proofT}
Let us  assume to the contrary, namely, there is a symmetric deadlock-free
mutual exclusion algorithm for $n$ processes using $m\geq 1$ anonymous
RMW registers such that for some positive integer $1<\ell\leq n$, $m$
and $\ell$ are not relatively prime.  This means that there is a
number $1<\ell\leq m$ such that $\ell$ divides $m$.

Let us  arrange the $m$ RMW registers on a ring with $m$ nodes where each
register is placed on a different node.  Then, let us  pick $\ell$
processes. For simplicity let us call these processes
$p_0,...,p_{\ell-1}$.
To each one of the $\ell$ processes, we assign an initial RMW
register (namely, the first register that the process accesses) such that
for every two processes $p_i$ and $p_{i+1~(mod~\ell)}$, the distance
between their initial registers is exactly $m/\ell$ when walking on the
ring in a clockwise direction. Here we use the assumption that $\ell$
divides $m$.

The lack of global names, allows us to assign for
each process an initial RMW register and an ordering of the registers
which determines how the process scans the registers.

An execution in which the $\ell$ processes are running in \emph{lock
  steps}, is an execution where we let each process takes one step (in
the order $p_0,...,p_{\ell -1}$), and then let each process takes
another step, and so on.
For process $p_i$ and integer $k$, let $order(p_i, k)$ denotes the
$k^{th}$ new register that $p_i$ accesses during an execution where
the $\ell$ processes are running in lock steps, and assume that we
arrange that $order(p_i, k)$ is the register whose its distance from $p_i$'s
initial registers is exactly $(k-1)$, when walking on the ring in a
clockwise direction.

We notice that $order(p_i, 1)$ is $p_i$'s initial register,
$order(p_i, 2)$ is the next new register that $p_i$ accesses and so
on.  That is, $p_i$ does not access $order(p_i, k+1)$ before accessing
$order(p_i, k)$ at least once, but for every $j\leq k$, $p_i$ may
access $order(p_i, j)$ several times before accessing $order(p_i, k+1)$
for the first time.

With this arrangement of RMW registers,  we run the $\ell$
processes in lock steps. Since only comparisons for equality are
allowed, and all registers are initialized to a the same value
--which (to preserve anonymity) is not a process identity--
processes that take the same number of steps will be at the
same state, and thus it is not possible to break symmetry.  It follows that
either all the processes will enter their critical sections at the
same time, violating mutual exclusion, or no process will ever enter
its critical section,  violating deadlock-freedom.  A contradiction.
\renewcommand{\toto}{theo-bound-RMW}
\end{proofT}

\section{Conclusion}
\label{sec:conclusion}
``Anonymous shared  memory'' means  there is no a priori
agreement among the  processes  on the names of the shared registers.
Moreover,  ``symmetric algorithm'' means that
the process identities define a specific data type
with no internal structure (such as a total order) and  no relation
with other data type (hence an identity cannot be compared with an integer).
Identities can only  be read, written, and compared with equality.

Considering two types of  anonymous registers,
namely atomic read/write (RW) registers and
atomic read/modify/write (RMW) registers,
This paper presented several results on symmetric mutual exclusion
algorithms,  summarized in Table~\ref{table:memory-anonymous-summary}.

\begin{table}[h!]
\begin{center}
  \begin{tabular}{|c|c|c|}
    \hline
     Registers    & RW anonymous    &   RMW anonymous    \\
    \hline
    \hline
    ~Sufficient condition~(algorithm)~ &  This paper     & This paper  \\
    \hline
    ~Necessary condition~  &  \cite{T17}\tablefootnote{Notice that the lower bound for the RW model from \cite{T17}, follows immediately from our stronger lower bound for the RMW model present in this paper.}  &  This paper \\
    \hline
  \end{tabular}

\end{center}
\caption{A global picture for $n$-process anonymous mutex ($n\geq 2$)}
\label{table:memory-anonymous-summary}
\end{table}

The symmetric deadlock-free mutex algorithm built on top of an
anonymous memory of $m$ atomic read/write registers works for $m\in
M(n)\setminus\{1\}$, where $M(n)=\{m: \forall~ \ell: 1<\ell \leq
n:~\ggcd(\ell,m)=1\}$, while the algorithm for $m$ atomic
read/modify/write registers works for $m\in M(n)$.  The necessity of
the first condition was proved in~\cite{T17}, while the necessity of
the second condition was proved in this paper.  The existence of the
algorithms presented in the paper proves these conditions are also
sufficient.

Let us remark that a system composed of a single anonymous
register is no really anonymous.  Hence, if we eliminate the
``pathological'' case $m=1$, $m\in M(n)$ is a necessary and sufficient
condition for symmetric deadlock-free mutex in both the read/write and
the read/modify/write anonymous register models. This shows a
fundamental computability difference separating the ``memory
anonymity'' adversary (which operates before the execution and is
consequently {\it static}) and the ``process crash'' adversary (which
operates during the execution and is consequently {\it dynamic}), for
which read/write and read/modify/write registers (such as
compare\&swap) are located at the two extremes of the synchronization
power hierarchy  as defined in~\cite{H91}.  (Let us remind that mutex
can be solved neither in the read/write nor in  the read/modify/write
non-anonymous register models in the presence of process crashes.)
Last but not least, a noteworthy property of the two algorithms that
have been presented lies in their simplicity.

\section*{Acknowledgements}
  Zahra Aghazadeh and Philipp Woelfel were partially
  supported by the Canada Research Chairs program and by the Discovery
  Grants program of the Natural Sciences and Engineering Research
  Council of Canada (NSERC).  Michel Raynal was partially supported by
  the French ANR project 16-CE40-0023-03 DESCARTES devoted to layered
  and modular structures in distributed computing.

\end{document}